%% file: ms.tex
\documentclass[twocolumn]{aastex63}

\usepackage{xcolor}

\usepackage{bm}

\usepackage[T1]{fontenc}
\usepackage{ae,aecompl}
\usepackage{times}

\usepackage{mathtools}

\usepackage{multirow}

\usepackage{newtxtext,newtxmath}

\usepackage{xspace}
\usepackage{ulem}

\newcommand{\unitflux}{\,erg\,cm$^{-2}$\,s$^{-1}$}
\newcommand{\unitlumi}{\,erg\,s$^{-1}$}
\newcommand{\mission}[1]{\textit{#1}}
\newcommand{\msun}{\mathrm{M}_\odot}
\newcommand{\mbh}{M_\mathrm{BH}}

\newcommand{\ledd}{L_\mathrm{Edd}}

\received{June 1, 2019}
\revised{January 10, 2019}
\accepted{\today}
\submitjournal{ApJ}

\shorttitle{TDE in AGN J0227-0420}
\shortauthors{Liu et al. 2019}
\begin{document}

\title{A Tidal Disruption Event Candidate Discovered in the Active Galactic Nucleus SDSS J022700.77-042020.6}

\correspondingauthor{Zhu Liu}
\email{liuzhu@nao.cas.cn}

\author{Zhu Liu}
\affiliation{Key Laboratory of Space Astronomy and Technology, National Astronomical Observatories, Chinese Academy of Sciences, Beijing 100101, China}

\author{Dongyue Li}
\affiliation{Key Laboratory of Space Astronomy and Technology, National Astronomical Observatories, Chinese Academy of Sciences, Beijing 100101, China}
\affiliation{University of Chinese Academy of Sciences, School of Astronomy and Space Science, Beijing 100049, China}

\author{He-Yang Liu}
\affiliation{Key Laboratory of Space Astronomy and Technology, National Astronomical Observatories, Chinese Academy of Sciences, Beijing 100101, China}
\affiliation{University of Chinese Academy of Sciences, School of Astronomy and Space Science, Beijing 100049, China}

\author{Youjun Lu}
\affiliation{National Astronomical Observatories, Chinese Academy of Sciences, Beijing 100101, China}
\affiliation{University of Chinese Academy of Sciences, School of Astronomy and Space Science, Beijing 100049, China}

\author{Weimin Yuan}
\affiliation{Key Laboratory of Space Astronomy and Technology, National Astronomical Observatories, Chinese Academy of Sciences, Beijing 100101, China}
\affiliation{University of Chinese Academy of Sciences, School of Astronomy and Space Science, Beijing 100049, China}

\author{Liming Dou}
\affiliation{Center for Astrophysics, Guangzhou University, Guangzhou 510006, China}

\author{Rong-Feng Shen}
\affiliation{School of Physics and Astronomy, Sun Yat-Sen University, Zhuhai, Guangdong 519082, China}

%% Note that the \and command from previous versions of AASTeX is now
%% depreciated in this version as it is no longer necessary. AASTeX 
%% automatically takes care of all commas and "and"s between authors names.

%% AASTeX 6.3 has the new \collaboration and \nocollaboration commands to
%% provide the collaboration status of a group of authors. These commands 
%% can be used either before or after the list of corresponding authors. The
%% argument for \collaboration is the collaboration identifier. Authors are
%% encouraged to surround collaboration identifiers with ()s. The 
%% \nocollaboration command takes no argument and exists to indicate that
%% the nearby authors are not part of surrounding collaborations.

%% Mark off the abstract in the ``abstract'' environment. 

\begin{abstract}

	We report the discovery of a Tidal Disruption Event (TDE) candidate occurring in the Active Galactic Nucleus SDSS J022700.77-042020.6. A sudden increase in flux of J0227-0420 during the second half of 2009 is clearly shown in the long-term optical, UV, and NIR light curves. A plateau phase, following an initial decline, is seen in the NUV and optical \textit{u, g, r, i} light curves. Moreover, we find possible evidence that the plateau phase in the NUV band may lag behind the optical ones by approximately 70--80\,days with also a much shorter duration, i.e. $\sim$7--15\,days against $\sim$40--50\,days. The long-term NUV/optical (after the plateau phase), NIR and MIR light curves can be well characterized with a form of $L(t)\propto t^{-\beta}$, consistent with the expectation of a TDE. The plateaus can be explained if the stellar streams collide with the pre-existing AGN disk at different radii. Though the overall fallback rate decreases, the material in the outer disk gradually drifts inward and increases the local accretion rate at the inner region, producing the optical and UV plateaus. The possible lag between the optical and NUV plateaus can then be attributed to viscosity delay. The index $\beta$ of the NIR $J, H, K_s$ bands ($\sim1.4-3.3$) is steeper than that of the UV/optical ($\sim0.7-1.3$) and MIR bands ($\sim0.9-1.8$), which may suggest that a certain fraction of the dust in the inner region of the dusty torus may be sublimated during the TDE phase. Our results indicate that, due to collisions between stellar debris and pre-existing disk, the light curves of TDEs occurring in AGN may show distinctive features, which may shed new light on the accretion process.
	
\end{abstract}
%% Keywords should appear after the \end{abstract} command. 
%% See the online documentation for the full list of available subject
%% keywords and the rules for their use.

\section{Introduction}
A star of mass $M_{\star}$ and radius $r_{\star}$ will be tidally disrupted if it enters the tidal radius, $r_t \sim r_{\star}(M_\mathrm{BH}/M_{\star})^{1/3}$, of a black hole (BH). About half of the disrupted debris will then fall back onto the BH, forming an accretion disk and producing intensive radiation \citep{rees_1988, evans_kochanek1989}. Such events are called as Tidal Disruption Events (TDEs, see \citealt{komossa_2015} for a recent review). The peak luminosity of TDE can range from a few per cent to close to the Eddington luminosity ($\ledd=1.38\times10^{38}\mbh/\msun$\unitlumi, \citealt{hung_etal2017, wevers_etal2019a}), thus an otherwise quiescent galaxy can be as bright as the Active Galactic Nuclei (AGNs). TDEs can yield important insight in the accretion physics as well as the formation of the accretion disk. It also provides an important tool to discover dormant BH, which is crucial for studying the BH demography.

Theoretical calculation suggests that the radiation of TDEs will peak in the soft X-ray/UV bands \citep{rees_1988, evans_kochanek1989}. The luminosity of TDEs, after the disruption time $t_0$, will approximately evolve following a power-law, $L(t)\propto(t-t_0)^{-\beta}$ \citep{phinney1989, evans_kochanek1989}. The value of $\beta$ is likely wavelength dependent, e.g., $\beta\sim5/3$ in the X-ray band, while a flatter power-law with $\beta\sim5/12$ is expected in the UV and optical bands \citep{lodato_rossi2011}.

The first TDE candidate was discovered in the soft X-ray band using archival \mission{ROSAT} data \citep[][NGC 5905]{komossa_bade1999}. Thereafter, more candidates have been found in the X-ray band from \mission{ROSAT} archival data \citep[e.g.][]{komossa_greiner1999, grupe_etal1999, greiner_etal2000, khabibullin_sazonov2014}, and using \mission{XMM--Newton}, \mission{Chandra} and \mission{Swift/XRT} data (e.g. \citealt{cappelluti_etal2009, esquej_etal2008, lin_etal2011, saxton_etal2012, maksym_etal2014, saxton_etal2017}, see \citealt{komossa_2015} for a recent review). Benefitting from the launching and construction of UV and optical telescopes with relatively large field-of-view (FoV), many TDEs has been discovered in the UV and optical bands. For instance, three TDE candidates were reported in the UV band using \mission{GALEX} data \citep{gezari_etal2006, gezari_etal2008, gezari_etal2009}. Several dozens of TDEs were discovered in the SDSS, PTF, ASAS-SN and PanSTAARS survey in optical bands \citep[e.g.][]{komossa_etal2008, vanvelzen_etal2011, wang_etal2011, cenko_etal2012, holoien_etal2014, arcavi_etal2014, chornock_etal2014}. The photons emitted from TDEs may be reprocessed by the surrounding medium, producing an echo of the main event with a decay time that depends on the distance between the central BH and the surrounding medium. Such phenomenon has been observed in the optical spectrum and mid-Infrared (MIR) light curve of some TDEs \citep[e.g.][]{komossa_etal2012, wang_etal2012, dou_etal2016, dou_etal2017, jiang_etal2016, jiang_etal2017, vanvelzen_etal2016}.

Detection and confirmation of TDEs that take place in AGNs are more difficult than those in quiescent galaxies, since AGNs themselves are quite bright in almost the entire electromagnetic spectrum \citep{elvis_etal1994} and frequently show flare-like features \citep[e.g.][]{graham_etal2017}. Nevertheless, TDEs could occur in AGNs. The expected event rate in a gaseous disk with BH mass less than $10^7\msun$ is actually even higher than that in a quiescent galaxy \citep{karas_subr2007}. So far, only a few TDEs happened in AGNs are reported. \citet{blanchard_etal2017} reported the detection of a TDE (PS16dtm) in a narrow-line Seyfert 1 galaxy (NLS1) using multi-band data. The TDE nature of this source was further supported by the discovery of the MIR echo of the TDE from the dusty torus of the AGN \citep[][see also \citealt{yang_etal2019} for a MIR flare found in a Type II AGN]{jiang_etal2017}. \citet{yan_xie2018} suggested that disruption of a main sequence star is favored to explain the decades long exponential decay of the X-ray light curve in the low luminosity AGN NGC 7213.  Similar characteristic light curve was also found in the AGN GSN 069 which showed a super soft X-ray spectrum. \citet{shu_etal2018} proposed that the X-ray light curve and X-ray spectral properties of GSN 069 can be well explained by a TDE\footnote{X-ray quasi-periodic eruptions (QPEs) has recently been discovered in this source \citep{miniutti_etal2019}. The physical origin of the QPE is still not clear. The general property of GSN 069 is consistent with the expectation from TDE, but X-ray variability from AGN can not be ruled out.}. By analyzing the mid-IR data, \citet{jiang_etal2019} suggested that the energetic transient event in the active galaxy PS1-10adi \citep{kankare_etal2017} is likely caused by a TDE. The most distant TDE candidate, at redshift of $z\sim2.359$, was found in a quasar by studying the abundance ratio variability \citep{liu_dittmannetal2018}.

The total number of TDEs with well sampled light curve and spectroscopic data is still small (e.g. see \citealt{auchettl_etl2017} for the study of X-ray property of a sample of X-ray detected TDEs). Despite this the long-term light curve of those TDEs found in AGNs seems to show distinctive characteristics from TDEs happened in quiescent galaxies. For instance, the two TDEs found in NGC 7213 and GSN 069 showed a decay lasting for decades, which is rare in normal TDEs (but see \citealt{lin_etal2017}). A plateau phase lasting $\sim100$ days after reaching the peak luminosity is shown in the multi-band light curves of PS16dtm \citep{blanchard_etal2017}. Those observational results may suggest that the accretion process for TDEs happened in AGNs differs from that of normal TDEs (we note that the plateau phase has also been reported in some normal TDEs, see e.g. \citealt{wevers_etal2019, vanvelzen_etal2019}), probably due to the interaction of the stellar debris with the pre-existing accretion disk. \citet{kathirgamaraju_etal2017} suggested that the fallback of stellar debris can be stalled because of the interaction with the pre-existing accretion disk, leading to an abrupt cut-off in the light curve. Numerical simulation carried out by \citet{chan_etal2019} showed that the rapid inflow induced by the shocks, which are excited by the collision between the stellar debris and the pre-existing disk, may cause the interior disk to be depleted within a fraction of the mass return time. Part of the fallback material may eventually hit the disk at larger radii if the stream is heavy enough to penetrate the disk.

TDEs can potentially explain some of the observed phenomena in AGNs. For instance, \citet{merloni_etal2015} showed that a star tidally disrupted by a BH could explain the changing-look behavior of SDSS J015957.64+003310.5. Such events can also explain the unusually high nitrogen-to-carbon ratio in a small fraction of quasars \citep[e.g.][]{liu_etal2018}. Moreover, it is now generally believed that the growth of BH and the evolution of the BH spin are mainly contributed from major merger and accretion (the AGN phase). Theoretical calculation suggests that TDEs may dominate or significantly contribute to the mass growth for $\mbh\lesssim 10^6\msun$ \citep{milosavljevic_etal2006}. \citet{zhang_etal2019} showed that TDEs may dramatically change the spin distribution for BHs with mass $\sim10^6\msun$. Thus the contribution from TDEs should be taken into account for studies of mass growth, accretion history, and spin distribution of AGNs with low BH mass. In addition, the stellar debris of TDEs can possibly interact with the pre-existing accretion disk of AGNs. The debris may be able to change the accretion rate of a AGN dramatically (with peak accretion rate close to the Eddington accretion rate), providing an ideal laboratory to test the accretion theory, investigate the transition of accretion mode, and study the disk structure.

\begin{figure*}
	\includegraphics[width=1.0\textwidth]{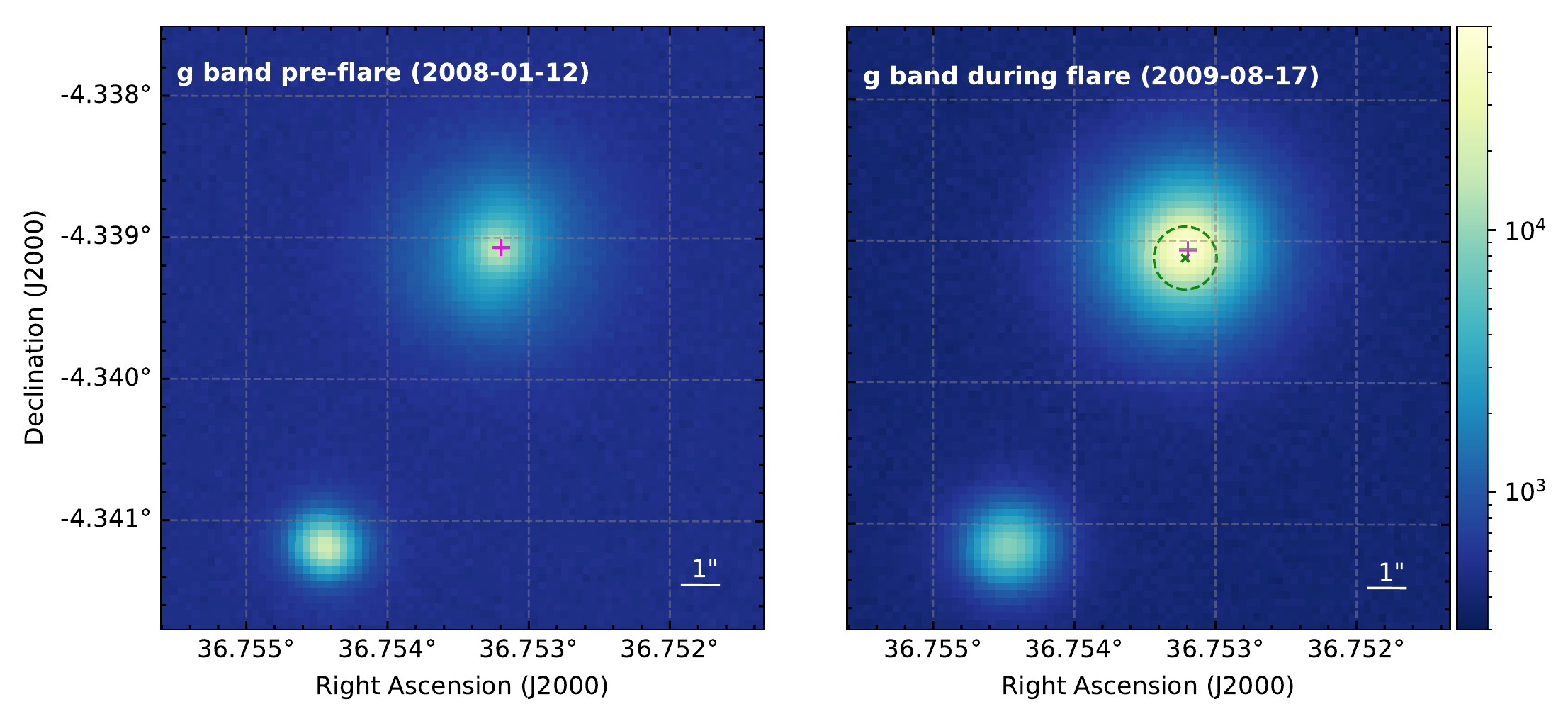}
	\vspace*{-6mm}
	\caption{\label{fig:flare_image}Left panel: one CFHT \textit{g} band image taken before the flare. The magenta plus marks the average position of the AGN from the observations before the flare; Right panel: one CFHT \textit{g} band image observed during the flare. The average position of the flare, measured from observations carried out after the flare, is marked with magenta plus. The green cross and the dashed circle represent the position of the AGN and its 90 per cent uncertainty given by \mission{Chandra} observation, respectively. Note that the position of the AGN, estimated from the \textit{g} band observations, is shown as green plus. However, the difference is too small to be clearly shown in this plot.}
\end{figure*}

In this work, we report the discovery of a new TDE candidate occurring in AGN SDSS J022700.77-042020.6 (hereafter J0227-0420). The AGN nature of J0227-0420 is revealed by the X-ray spectral property and the detection of the broad optical emission lines before the flare phase. The NUV, optical and infrared flux increased by a factor of $\gtrsim 7$ at around August 2009. The long-term multi-band light curves can be well described with the form of $\nu L_\nu(t)\propto(t-t_0)^{-\beta}$ (see Section\,\ref{subsec:modelling_lc}), suggesting that the dramatical increase in flux is due to a TDE happened at the center of the AGN. This paper is structured as follows: in Section\,\ref{sec:multi_data}, we described the multi-band data analysis; Evidence from optical spectroscopic and X-ray data analysis that revealed the AGN nature of J0227-0420 was presented in Section\,\ref{sec:bh_agn}; In Section\,\ref{sec:multi_lc}, we modeled the multi-band long-term light-curve of J0227-0420. We discussed and summarized our results in Section\,\ref{sec:discussion} and \ref{sec:summary}, respectively. Throughout this paper, we adopted a flat $\Lambda\mathrm{CDM}$ cosmological model with $H_0=69.3\,\mathrm{km\,s^{-1}}$, $\Omega_m=0.29$ and $\Omega_\Lambda=0.71$ \citep{hinshaw_etal2013}. All the quoted uncertainties correspond to the 90 per cent confidence level for one interesting parameter, unless specified otherwise.

\section{Multiband Data analysis\label{sec:multi_data}}
The sky region around the position of J0227-0420 has been extensively observed by several surveys in the past decades, e.g. the XMM-XXL survey \citep{menzel_etal2016} in X-ray band; the Deep Field of the Canada-France-Hawaii Telescope Legacy Survey (CFHTLS\footnote{\url{https://www.cfht.hawaii.edu/Science/CFHTLS}}) in optical band; the VISTA Deep Extragalactic Observations Survey (VIDEO, \citealt{jarvis_etal2013}), the VISTA Hemisphere Survey (VHS), and the VISTA Kilo-Degree Infrared Galaxy Survey (VIKING, \citealt{edge_etal2013}) in near-IR band; the UKIRT Deep Extragalactic Survey (DXS) in near-IR band, and the GALEX time domain survey \citep{gezari_etal2013} in UV band. An optical spectroscopy was taken with SDSS in 2012. In this section we describe the multi-band data analysis of J0227-0420.

\subsection{\mission{XMM-Newton}\label{subsec:multi_xmm}}
J0227-0420 was observed six times by \mission{XMM-Newton} with two observations carried out in 2002 and the rest four performed in 2017 (see Table\,\ref{tab:xray_fitting}). It is included in the AGN catalogue of the XMM-XXL Northern Sky Survey \citep{liu_etal2016}. Here we re-reduced all the \mission{XMM-Newton} data using the latest calibration files. The Observation Data Files (ODFs), retrieving from the \mission{XMM-Newton} Science Archive, were reduced using the \mission{XMM-Newtion} Science Analysis System software (\textsc{SAS}, version 16.1, \citealt{gabriel_etal2004}). For each observation, we ran the \textsc{SAS} tasks \textsc{emchain} and \textsc{epchain} to generate the event lists for the European Photon Imaging Camera (EPIC) MOS \citep{turner_etal2001} and pn \citep{struder_etal2001} detectors, respectively. High background flaring periods were identified and filtered from the event lists. For all the observations, a circular region with radius of $40$ and $45$\,arcsec was selected as the source region for the MOS and pn images, respectively. For the background region, a circular region, centered at the source position, with radius of $100$\,arcsec was chosen for the MOS cameras. While the background region of the pn camera was selected from a circular region centered at the same CCD read-out column as the source position. X-ray events with pattern $\leq12$ for MOS and $\leq4$ for pn were selected to extract the X-ray spectra. We used the \textsc{SAS} task \textsc{RMFGEN} and \textsc{ARFGEN} to generate the response matrix and the ancillary files, respectively. 

\subsection{\mission{Chandra}}
 \mission{Chandra} observed J0227-0420 on 2016 October 5 (13.4\,ks, ObsID: 18256) with the Advanced CCD Imaging Spectrometer (ACIS). \mission{Chandra} data were downloaded from the \mission{Chandra} data archive, and were reduced with \textsc{CIAO} (\citealt{fruscione_etal06}, ver 4.10) software package and calibration files CALDB (ver 4.7.6). The \textsc{chandra\_repro} script was used to reprocess the data. We ran \textsc{wavdetect} tool on the reprocessed \mission{Chandra} data to generate a source list, which also gave the coordinates of the J0227-0420 with $(\mathrm{RA, Dec}) = \mathrm{(02^h27^m0^s.77, -04^\circ20'20.83'')}$. In principle, the absolute astrometry of \mission{Chandra} detected sources can be improved by cross-matching \mission{Chandra} sources with the other source catalogue (e.g. \mission{GAIA}). However, due to the lack of bright X-ray sources in the field-of-view, this procedure does not improve the position uncertainty of J0227-0420. Thus the overall 90 per cent absolute astrometry uncertainty of \mission{Chandra} ($\sim0.8$\,arcsec\footnote{\url{http://cxc.cfa.harvard.edu/cal/ASPECT/celmon}}) is used in this work.

\subsection{\mission{GALEX}\label{subsec:multi_galex}}
The \mission{GALEX} satellite frequently observed the sky region around J0227-0420 in the UV band. In total, there are 178 observations with an accumulative exposure time of 233 ks in NUV band, and 124 observations with 163 ks total exposure time in FUV band. The \textsc{gPhton} \citep{million_etal2016b, million_etal2016a} Python package\footnote{\url{https://archive.stsci.edu/prepds/gphoton}} was used to analyze the \mission{GALEX} NUV (1771--2831\AA) and FUV (1344--1786\AA) data of J0227-0420. The intensity maps in the NUV and FUV bands were generated using the \textsc{gMap} task. To measure the UV photometry, a circular region with radius of 8\,arcsec was selected as the source region for both the NUV and FUV data, while an annulus (concentred with source) with inner radius of 80\,arcsec and outer radius of 88\,arcsec was selected as the background region. The \textsc{gAperture} task was then used to calculate the background subtracted AB magnitude and flux for each individual observation.

\subsection{\mission{CFHT}\label{subsec:multi_optical}}
J0227-0420 locates in the region covered by the Deep Field 1 of the CHFTLS, and was routinely observed in $u, g, r, i, z$ bands with the MegaCam camera on the $3.6\,\mathrm{m}$ \mission{CFHT} telescope. We searched archival data in the \mission{CFHT} Science Archive within a radius of 30\,arcsec (account for the fact that the astrometry is not well corrected in the calibrated image) at the source position, which resulted in 4388 observations in total in the 5 bands. Due to strong contamination from the host galaxy, the \textit{z} band data were not used in this work. All the observations in the $u, g, r, i$ bands carried out after 2009 were selected, while only those have longer exposure time were chosen for observations performed from 2005 to 2009. In addition, filters on the MegaCam have been changed or upgrade several times since 2004. To avoid potential uncertainties caused by the differences in the filters, only observations with the following filters were chosen: U9301, G9401, R9601, and I9702\footnote{Note that the original I band filter (I9701) was broken in Oct 2007, and was replaced by the I9702 filter. So our \textit{I} band data only covered the time period from 2007 October to 2014 September.}.

The calibrated multi-band image data, which were reduced using the Elixir pipeline \citep[version 2,][]{magnier_cuillandre2004}, were downloaded from the Canadian Astronomy Data Centre (CADC)\footnote{\url{http://www.cadc-ccda.hia-iha.nrc-cnrc.gc.ca/en/cfht}}. The astrometric corrections were performed using the \textsc{Scamp} software\footnote{\url{https://github.com/astromatic/scamp}} \citep[version 2.7.1]{bertin_2006}. The position of J0227-0420, after astrometric correction, was marked as magenta plus in the two \textit{g} band images observed before (left panel of Figure\,\ref{fig:flare_image}) and after (right panel Figure\,\ref{fig:flare_image}) the flare. Photometric measurements were done with the \textsc{SExtractor} software\footnote{\url{https://github.com/astromatic/sextractor}} (\citealt{bertin_arnouts1996}, version 2.25.0). To avoid contamination from a nearby star, a circle with radius of 5\,arcsec was chosen to measure the magnitude and flux for each individual observation. Observations of which the source is saturated, i.e., a non-zero flag parameter was reported by \textsc{SExtractor}, were excluded. In addition, we also excluded observations of which the source is located on bad columns on the CCD chip. To check our photometric measurements, we compared our measured magnitudes of some stars with the results from SDSS. The measured \mission{CFHT} magnitudes were first converted to SDSS magnitudes using the color conversion from the CADC website\footnote{\url{http://www.cadc-ccda.hia-iha.nrc-cnrc.gc.ca/en/megapipe/docs/filt.html}}. We found that our measurements of the $g, r, i$ magnitudes matched well with the SDSS results (within $\pm0.02$\,mag), while a large offset of $\sim0.3$\,mag was found for the $u$ band measurements. This is consistent with previous study\footnote{\url{http://www.cfht.hawaii.edu/Instruments/Imaging/MegaPrime/PDFs/megacam.pdf}} (though we gave a relatively larger offset, i.e. $\sim0.3$ vs. $\sim0.15$\,mag). To roughly correct the $u$ band photometry, all the measured \textit{u} band magnitudes were subtracted by 0.3\,mag. We then converted the instrumental magnitudes to AB magnitudes.

\subsection{\mission{VISTA}\label{subsec:multi_nir}}
The \mission{VISTA} \citep{emerson_etal2006, dalton_etal2006} is a 4-m class wide field survey telescope equipped with a near infrared camera, VIRCAM, which has five broad band filters ($Y, Z, J, H, K_\mathrm{s}$) and a narrow band filter at $1.18\,\mathrm{\mu m}$. J0227-0420 is within the sky region covered by several NIR surveys in $Y, Z, J, H, K_\mathrm{s}$ bands with the VISTA telescope, e.g. the VHS, VIKING and VIDEO surveys, from 2009 November till the end of 2013. In this work, the photometric measurements were obtained from the VIDEO DR5\footnote{\url{http://wfaudata.roe.ac.uk/videoDR5-dsa/TAP}}, VIKING DR4\footnote{\url{http://wfaudata.roe.ac.uk/vikingDR4-dsa/TAP}}, and VHS DR4\footnote{\url{http://wfaudata.roe.ac.uk/vhsDR4-dsa/TAP}} catalogues which were analyzed and compiled using the VISTA Data Flow System (VDFS) at the Cambridge Astronomical Survey Unit (CASU). Similar to the \mission{CFHT} \textit{z} band, the \textit{Z} band data were not used due to significant contamination from the host galaxy. To be consistent with the procedure adopted for the CFHT data, the photometric values measured from a 5\,arcsec circle were chosen. We converted the instrumental magnitudes to AB magnitudes (and then to flux) using the color transfer function presented in \citet{gonzalez-fernandez_etal2018}.

\subsection{\mission{WISE}\label{subsec:multi_wise}}
J227-0440 is also detected by the Wide-field Infrared Survey Explorer \citep[WISE,][]{wright_etal2010}, which is a satellite that surveys the sky in MIR bands. We searched the ALLWISE \citep{mainzer_etal2011} and the NEOWISE Reactivation data release catalogues \citep{mainzer_etal2014} at the source position of J0227-0420 with a matching radius of 1\,arcsec. This results in a total of 37 and 101 single exposure observations from the ALLWISE and NEOWISE, respectively. The W1 ($3.4\,\mathrm{\mu m}$) and W2 ($4.6\,\mathrm{\mu m}$) Vega magnitudes and uncertainties were obtained from the NASA/IPAC Infrared Science Archive (IRSA). The Vega magnitudes were then converted to the source flux density using the equations and parameters list on the WISE data processing website\footnote{\url{http://wise2.ipac.caltech.edu/docs/release/allsky/expsup/sec4_4h.html}}. We note that, due to the relatively large angular resolution of WISE (6.1 and 6.4 arcsec for W1 and W2, respectively), the WISE magnitudes are contaminated by nearby stars.

\section{The AGN nature and the BH mass of J0227-0420\label{sec:bh_agn}}
J0227-0420 is one of the targets in the XMM-XXL Northern Sky Survey that has optical follow-up observations with the SDSS/BOSS spectrograph. It was classified by \citet{menzel_etal2016} as an eAGN-SFG, i.e. a star-forming galaxy with high X-ray emission indicating accretion process at the center of the galaxy. Here we reanalyzed the X-ray spectra and the 2012 SDSS optical spectrum of J0227-0420.

\begin{figure*}
	\vspace*{-2cm}
	\includegraphics[width=\textwidth]{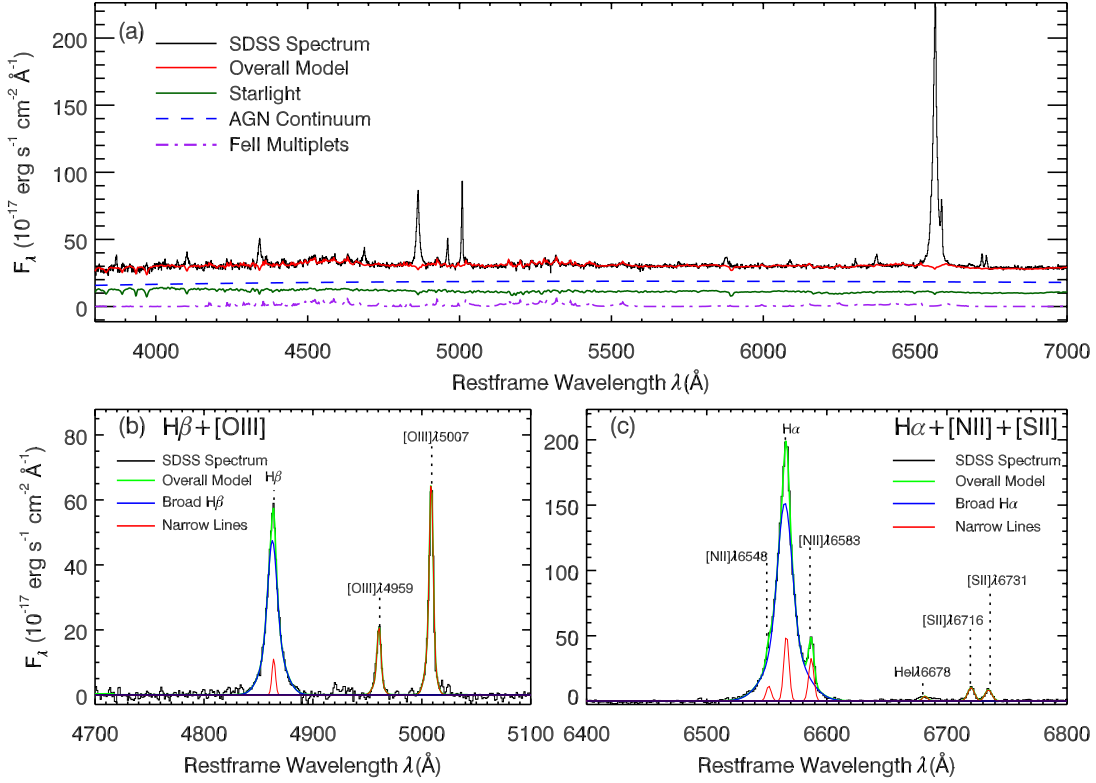}
	\vspace*{-11.0cm}
	\caption{\label{fig:sdss_spec}Upper panel: SDSS optical spectrum of J0227-0420(black), the total model (blue), the decomposed components of the host galaxy (red), the AGN continuum (green), and the \ion{Fe}{2} multiplets (purple); Bottom-left: emission-line profile fitting in the \ion{H}{2}+[\ion{O}{3}] region; Bottom-right: emission-line profile fitting in the ion{H}{1}+[\ion{N}{2}]+[\ion{S}{2}] region.}
\end{figure*}

\subsection{X-ray property of J0227-0420\label{subsec:xray_agn}}
The \textsc{Xspec} \citep[version 12.10][]{arnaud1996} software was used to analyse the X-ray spectra. Galactic absorption was always included (\textsc{TBabs} model in \textsc{Xspec}, abund is set to wilm), and the value ($2.32\times10^{20}\,\mathrm{cm^{-2}}$) was obtained using the \textsc{NH} tools in \textsc{HEASOFT}. No evidence for strong host galaxy absorption was found, i.e. only upper-limits that were smaller than the Galactic absorption can be given, thus we did not include any host galaxy absorption in our model. We noted that none of the \mission{XMM--Newton} spectra was taken during the flare phase.

\input{xray_spec_fit_info}

The X-ray spectra of J0227-0420 can be well fitted with a model consisting of a power-law plus a blackbody component. An emission line feature at around $6.4$\,keV, which should be the narrow Fe\,$\mathrm{K\alpha}$ commonly found in the X-ray spectra of AGNs, was also shown in the spectra with higher S/N. Thus an additional Gaussian component was added to model the high S/N spectra. The best-fitting parameters are listed in Table\,\ref{tab:xray_fitting}. The photon index of the power-law component as well as the temperature of the blackbody is consistent within uncertainty for the six observations, except for one observation (0780451601) of which only the MOS2 data were available. The best fitting photon index, $\Gamma\sim2.38$, is steeper than the average value in Seyfert galaxies, i.e. $\Gamma\sim1.6-2.1$ \citep{nandra_etal2007}, but still in agreement with the typical value found in the NLS1s. It is also clear (see Table\,\ref{tab:xray_fitting}) that J0227-0420 only shows mild X-ray variability with the 0.3--10\,keV X-ray luminosity in the range of $\sim8.5-12.0\times10^{42}\,$\unitlumi. The bolometric luminosity of the AGN, estimated from its 2--10\,keV X-ray luminosity, is $\sim8\times10^{43}\,$\unitlumi assuming a bolometric correction of 28 \citep{ho2008}. The X-ray spectral properties of J0227-0420, e.g. the photon index and temperature of the soft X-ray component, are typical of NLS1s. 

\subsection{Optical spectral analysis and the mass of the central BH\label{subsec:bh_mass}}

In this subsection, we summarized the main procedures to fit the continuum and emission line profiles of the SDSS spectrum. The details of the optical spectrum analysis can be found in \citet{dong_etal2012} and \citet{liu_etal2018}. Due to a relatively large fiber aperture of 2\,arcsec, the SDSS spectrum is significantly contaminated by the host galaxy starlight. In order to remove the starlight, six synthesized galaxy spectral templates \citep{lu_etal2006} were used to model the host galaxy starlight \citep{zhou_etal2006}. We used two separate sets of analytic templates \citep{veron-cetty_etal2004, dong_etal2008} to model the narrow-line and broad-line \ion{Fe}{2} emission, respectively. Following \citet{dong_etal2005}, all the other emission lines were fitted incrementally with as many Gaussians as statistically justified, e.g. a reduced $\chi^2$ less than 1.2.

The SDSS spectrum (black) as well as the best-fitting continuum model (blue) can be found in the upper panel of Figure\,\ref{fig:sdss_spec}. The decomposed host galaxy component is shown in red, while the AGN continuum is marked with green. The red and purple lines represent the contribution from the starlight and \ion{Fe}{2}, respectively. In the lower-left panel, the best-fitting \ion{H}{2} and [\ion{O}{3}] line profiles are shown, while in the lower-right panel, we show the best-fitting line profiles for the \ion{H}{1}, [\ion{N}{3}] and [\ion{S}{2}] emission lines.

The best-fitting FWHM of the narrow (e.g. [\ion{O}{3}] 5007) and broad (e.g. H$\alpha$) emission lines are $278\pm25$ and $775\pm37\,\mathrm{km\,s^{-1}}$, respectively. The small FWHM of the broad emission line suggests that J0227-0420 is likely to be a NLS1. The luminosity of the broad H$\alpha$ is $2.3\times10^{41}$\unitlumi. The BH mass of J0227-0420 is calculated to be $\sim10^6\msun$ (with typical uncertainty of $\sim0.3-0.5$\,dex, \citealt{liu_etal2018}) using the formalism presented by \citet{greene_ho2007} which makes use of the luminosity and FWHM of the broad H$\alpha$ line. Alternatively, we also estimated the BH mass using the $M-\sigma_\star$ relation \citep{ferrarese_merritt2000, gebhardt_etal2000}. The velocity dispersion estimated from the optical spectral analysis is $\sigma_\star=92\pm10\,\mathrm{km\,s^{-1}}$, indicating a BH mass of $2-6\times10^6\msun$ \citep{gultekin_etal2009} which is consistent with the results estimated from the broad emission line. Hereafter the BH mass of $10^6\msun$ measured from the broad emission line is used for J0227-0420. The bolometric luminosity is estimated to be $\sim6.6\times10^{43}\,$\unitlumi by assuming $L_\mathrm{bol}=9.8\lambda L_{\lambda 5100{\mathrm{\AA}}}$ \citep[][the $\lambda L_{\lambda 5100{\mathrm{\AA}}}$ is calculated from the H$\alpha$ luminosity, \citealt{greene_ho2005}]{mclure_dunlop2004}. This results in an Eddington ratio of $\sim0.6$ during the AGN phase for J0227-0420.

%%%%%%%%%%SECTION: PROPERTIES OF THE FLARE%%%%%%%%%%%%%%
\section{Properties of the flare\label{sec:multi_lc}}

\begin{figure*}
	\includegraphics[width=\textwidth]{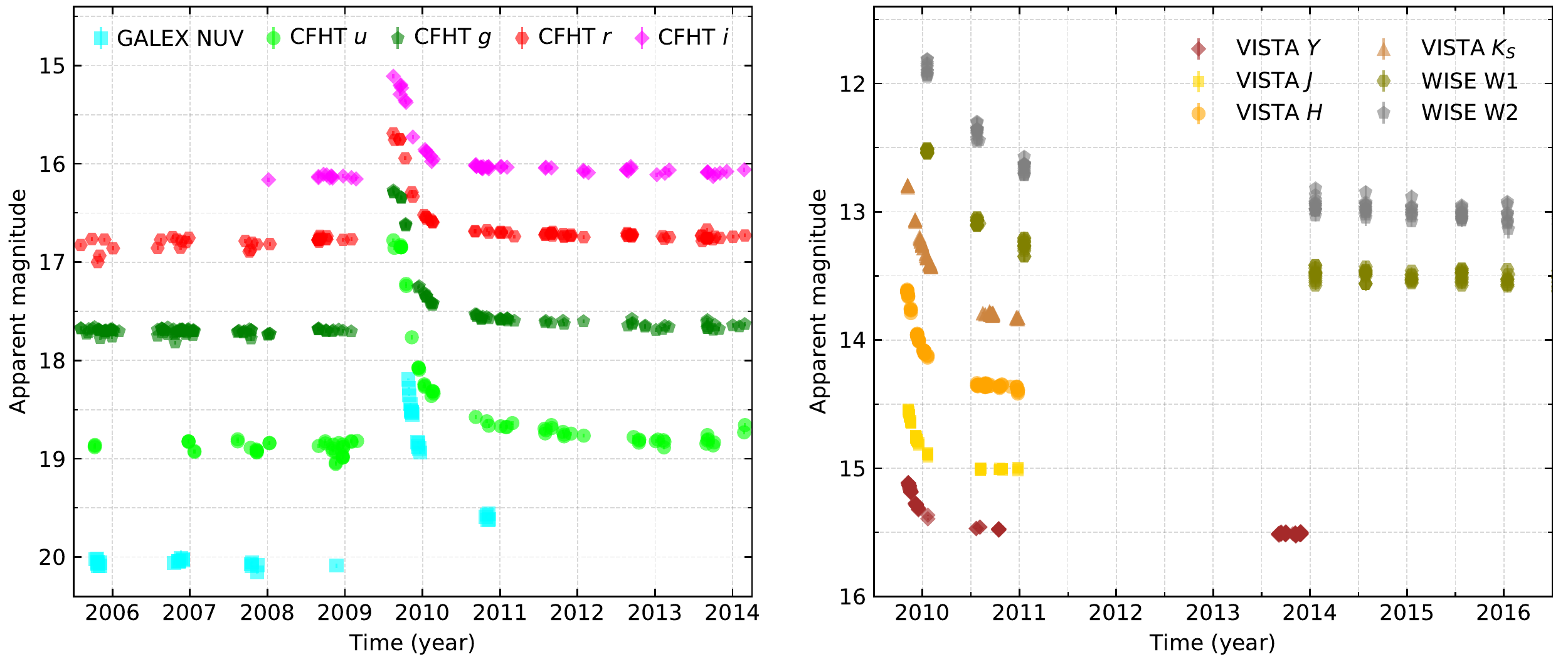}
	\vspace*{-6mm}
	\caption{\label{fig:multi_lc}Left-panel: long-term light curves measured from the CFHT $u$ (lime), $g$ (green), $r$ (red), $i$ (magenta) and \mission{GALEX} NUV bands (cyan); Right-panel: same as left panel, but for VISTA $Y$ (yellow), $J$ (gold), $H$ (orange), $K_\mathrm{s}$ (peru), as well as the WISE W1 (brown) and W2 (grey) bands. Note that the W1 and W2 magnitudes are in Vega magnitude, while all the others are in AB magnitude. For display purpose, the magnitude are shifted as following: \textit{u}+0.8, \textit{g}+0.2, \textit{i}-0.25, and NUV+0.75. }
\end{figure*}

The multi-band long-term light curves of J0227-0420 are shown in Figure\,\ref{fig:multi_lc}. In the vertical axises we show the apparent AB magnitude (except for the \mission{WISE} W1 and W2 data of which the Vega magnitudes are used) in different bands, while the decimal years are shown in the horizontal axises. A sudden increase in flux is clearly seen in the IR-to-NUV light curves, following by a long-term gradually decline lasting for more than 3 years. 

\subsection{Location of the flare}
The position of J0227-0420 of each \mission{CFHT} observation was given by \textsc{SExtractor}, after astrometric correction using \textsc{Scamp}. To better estimate the location of the flare, and to calculate the distance of the flare to the AGN, the mean of the measured Right Ascension (RA) and Declination (Dec) from all the \mission{CFHT} \textit{g} band observations before and after the flare is used as the location of the AGN and the flare, respectively. The $1\sigma$ position uncertainty is then estimated from the standard deviation of the measured RA and Dec values. This gives a position of $(\mathrm{RA, Dec}) = \mathrm{(02^h27^m0^s.78, -04^{\circ}20'20.66'')}$ for AGN, and $(\mathrm{RA, Dec}) = \mathrm{(02^h27^m0^s.77, -04^\circ20'20.63'')}$ for the flare (see Figure\,\ref{fig:flare_image}). The $1\sigma$ uncertainty is about $0.06\,\mathrm{arcsec}$ for both RA and Dec. The distance between the locations of the AGN and the flare, e.g. $\Delta\mathrm{RA}=0.07\,\mathrm{arcsec}$ and $\Delta\mathrm{Dec}=0.03\,\mathrm{arcsec}$, is comparable to (even less than) the position uncertainty, suggesting that the flare locates at the center of the galaxy, i.e. from the region consistent with the position of the AGN with a $1\sigma$ upper limit of $84\,\mathrm{pc}$.

In the right panel of Figure\,\ref{fig:flare_image}, the position measured from a \mission{Chandra} observation carried out in 2016 October is marked as green cross. The green dashed circle represents the typical $1\sigma$ astrometric uncertainty ($0.8\,\mathrm{arcsec}$) of \mission{Chandra}. It is clear that the position of the flare (as well as the AGN) measured from optical data is consistent with the value obtained from \mission{Chandra} X-ray measurement, i.e. position of the AGN.

\subsection{Light curve modeling\label{subsec:modelling_lc}}
\begin{figure*}
	\includegraphics[width=\textwidth]{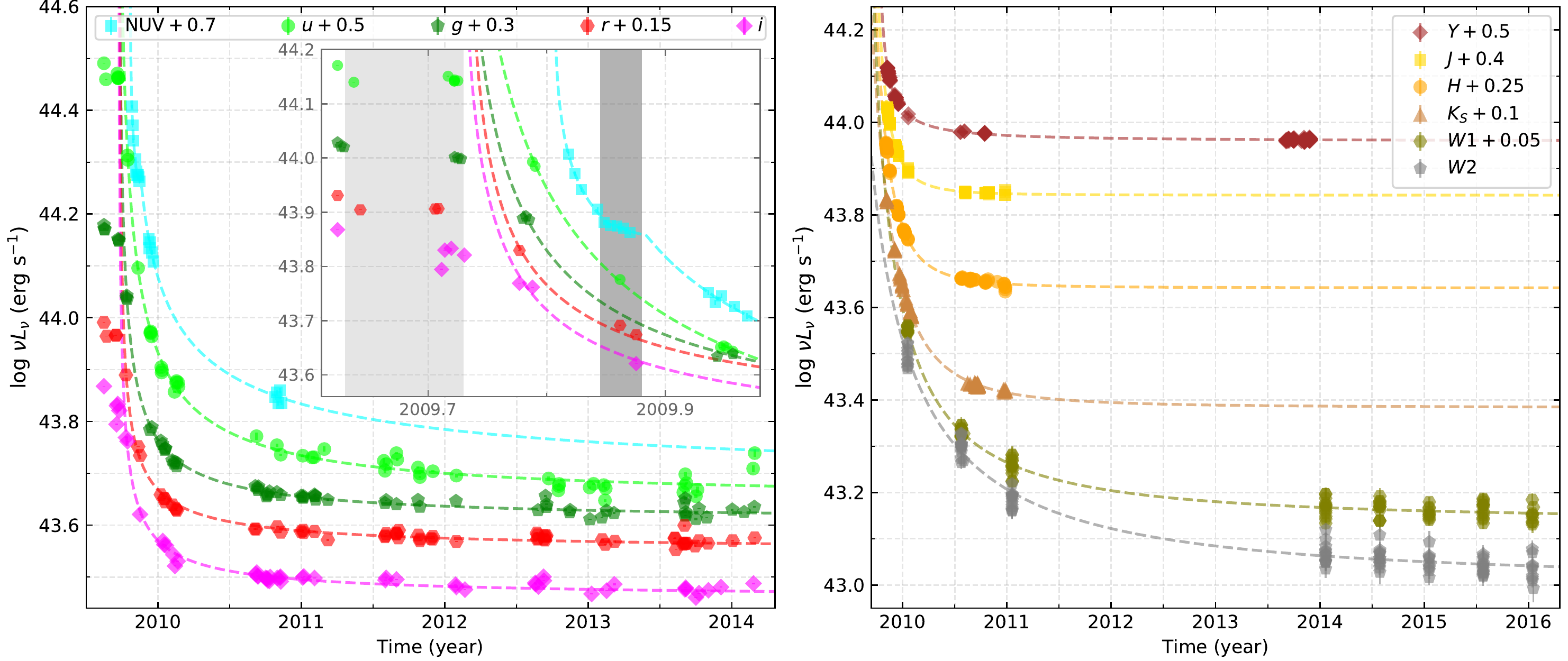}
	\vspace*{-6mm}
	\caption{\label{fig:opt_uv_lc}Left panel: evolution of the monochromatic luminosity (calculated from ) at the central wavelength of the NUV, \textit{u, g, r, i} bands. The inset shows a zoom-in view of the results taken before 2010. Plateau phases are clearly seen in the optical (gray shadow) and NUV (dark gray shadow) light curves. The dashed lines are calculated from the best-fitting models for the monochromatic flux with the form of $\nu f_\nu=A*(t-t_0)^{-\beta} + \mathrm{constant}$ (the function to model NUV data is different, see Section\,\ref{subsec:modelling_lc} for details); Right panel: the same as left panel, but for the \textit{Y, J, H, K$_S$, W1, and W2} bands. The color scheme is the same as Figure\,\ref{fig:multi_lc}. For display purpose, the monochromatic luminosity are shifted.}
\end{figure*}

To properly characterize the multi-band flux/luminosity evolution of the flare, the total contribution from the AGN and the host galaxy should be subtracted. For the \textit{Y, J, H, K}$_\mathrm{S}$ bands, the light curves were fitted with a model in the form of $\nu f_\nu=A*(t-t_0)^{-\beta} + \mathrm{constant}$, where the constant component was added to account for the contamination from the AGN and host galaxy, and $t_0$ represents the disruption time of the flare. While for the NUV and optical bands, the total contribution was estimated using the average flux measured before 2009 January (before the flare phase), as it is clearly from Figure\,\ref{fig:multi_lc} that the flux before 2009 January in these bands, which represents the total radiation from the AGN and host galaxy, did not vary significantly. In addition, a plateau is clearly shown in the light curves of the \textit{u, g, r, i} bands (the gray shadow area in the inset of Figure\,\ref{fig:opt_uv_lc}). We thus modeled the optical \textit{u, g, r, i} data (AGN and host galaxy subtracted) taken after 2009 August with a power-law form $\nu f_\nu = A * (t-t_0)^{-\beta}$.

We also tried to model the entire NUV light curve with the power-law form. However, a systematic excess respected to the best-fitting power-law at around the time duration 2009.8--2009.87 was found, indicating the presence of a plateau phase in the NUV band (the dark gray shadow area in the inset of Figure\,\ref{fig:opt_uv_lc}). Thus the NUV light curve was fitted with the following function (a broken power-law model):
$$ \nu f_\nu= \begin{dcases}
A*(t-t_0)^{-\beta_0} & t<t_0 \\[1.5mm]
A*\frac{(t_1 - t_0)^{\beta_1 - \beta_0}}{(t-t_0)^{\beta_1}}  & t_1<t<t_2 \\[1.5mm]
A*\frac{(t_1 - t_0)^{\beta_1 - \beta_0}}{(t-t_0)^{\beta_2}} (t_2 - t_0)^{\beta_2 - \beta_1}  & t>t_2
\end{dcases} $$
where $t_0$ is the disruption time. While $t_1$ and $t_2$ are the start and end time of the plateau phase, respectively.

\input{lc_fit_results}

A least-square optimization method was adopted to fit the multi-band light curves\footnote{We used the Python package \textsc{lmfit}. \url{https://lmfit.github.io/lmfit-py}}. The \textsc{emcee} Python package\footnote{\url{http://dfm.io/emcee}} \citep{foreman-mackey_etal2013} was used to obtain the best-fitting parameters, and to estimate the $1\sigma$ uncertainties. The best-fitting results are plotted in Figure\,\ref{fig:opt_uv_lc} (the monochromatic luminosities are calculated from the monochromatic fluxes), and given in Table\,\ref{tab:lc_fit}. As can be seen in Figure\,\ref{fig:opt_uv_lc}, the multi-band light curves can be well fitted with the power-law model (or multiple power-laws for NUV data) with the value of power-law index depending on the wave-band (though with large uncertainties), i.e. $\beta\sim0.73-1.29$ in the optical bands, $\sim0.99-3.28$ in the NIR bands, and $\sim0.88-1.77$ in the MIR bands. The best-fitting $t_{0}$ were also listed in Table\,\ref{tab:lc_fit}, however, we noted that those values are probably not the true disruption time because the peak flux were probably not observed, and that the plateau phase was not included in our modeling for the optical data. For the NUV data, the best-fitting indices are 0.29, 0.13 and 0.67 for $\beta_0$, $\beta_1$ and $\beta_2$, respectively. The best-fitting disruption time is $t_\mathrm{0}=2009.81$. The start and end times of the NUV plateau phase are $t_\mathrm{1}=2009.85$ and $t_\mathrm{2}=2009.88$, respectively.

To further tested whether the broken power-law model is indeed favored over the simple power-law model for NUV data, we performed model selection using both the Akaike Information Criterion \citep[AIC,][]{akaike1974} and the Bayesian Information Criterion \citep[BIC,][]{schwarz1978}. The broken power-law model gives smaller AIC and BIC values, suggesting that the model with a plateau phase is a better description of the NUV data. Similar tests were also carried out for the \textit{CFHT} optical data to check whether or not an optical plateau phase emerged at approximately the same time and duration of the NUV plateau phase. Due to the sparse of the optical data at the time of the NUV plateau phase, we jointly fitted \textit{u, g, r, i} datasets with both the power-law and the broken power-law models. Except for the normalization parameter, which is different for each dataset, all the other parameters in the models are assumed to be the same and are linked together. The $t_1$, $t_2$, and $\beta_1$ cannot be well constrained, thus are fixed at the best-fitting values obtained from the NUV modeling. We found that the broken power-law, which has more parameters, is not preferred as it gives higher AIC and BIC, indicating that an optical plateau was unlikely occurred at the time of the NUV plateau phase. The decrease in the \textit{r} band during the NUV plateau phase, which matches well with a simple power-law model rather than the break power-law, provides further evidence to support the conclusion.

 The location of the flare and the characteristic of the multi-band long-term light curves suggest that the sudden increase of the multi-band flux is likely caused by a TDE occurring in J0227-0420.

\subsection{Bolometric luminosity and temperature evolution}
\begin{figure}
	\includegraphics[width=\columnwidth]{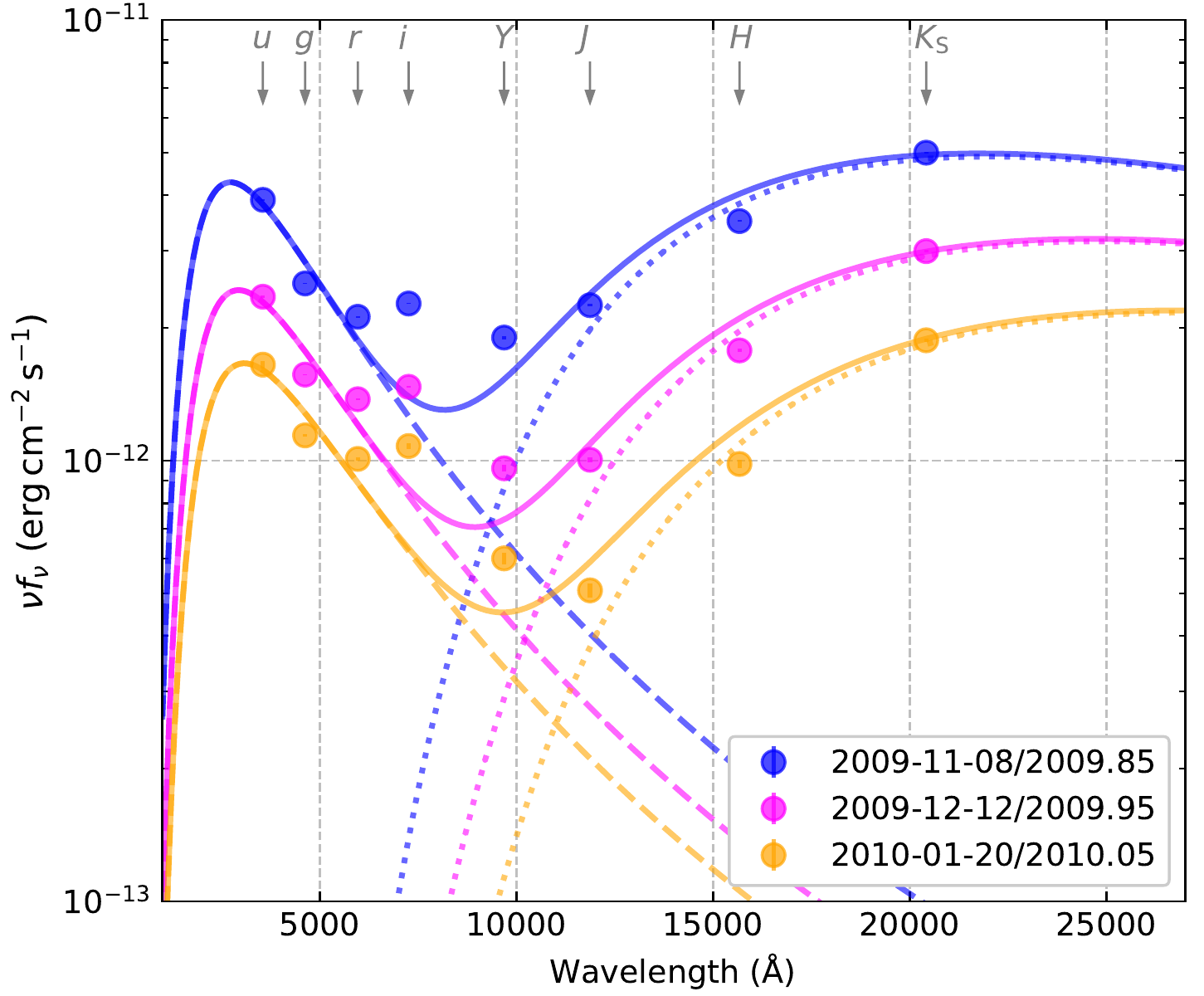}
	\caption{\label{fig:sed_fitting}The solid circles are the measured $\nu f_\nu$ for different bands at different observation dates. Note that the errors are too small to be seen in this figure. The dashed lines represent the best-fitting blackbody models for the optical \textit{u}, \textit{g}, and \textit{r} bands. While the best-fitting models for the NIR \textit{J}, \textit{H}, and $K_\mathrm{S}$ bands are shown as dotted lines. The solid lines are the summation of those two components. Different colors mark the observations at different epochs. Due to possible contamination from the host galaxy, the \textit{i} and \textit{Y} band data are not used in the modeling.}
\end{figure}

Previous studies found that the optical--UV radiation of most TDE can be fitted with a blackbody model with temperature of a few tens of thousands Kelvin \citep[e.g][]{holoien_etal2014, hung_etal2017, wevers_etal2017, vanvelzen_etal2019}. It is also known that the reprocess of the primary emission of TDEs by the surrounding dust will produce an echo in infrared \citep[e.g.][]{vanvelzen_etal2016, jiang_etal2016, jiang_etal2017}, which can also be modeled with a blackbody with temperature $\lesssim1500$\,K. In this work, we also modeled the SEDs of J0227-0420 with blackbody model at several epochs during which the observations at different bands were quasi-simultaneous. As an example, we show the optical to infrared broad-band photometric measurements for three epochs in Figure\,\ref{fig:sed_fitting}. It is clearly from Figure\,\ref{fig:sed_fitting} that two blackbody models are needed to fit the SEDs. Thus the optical bands (\textit{u}, \textit{g}, \textit{r}) and NIR bands ($J, H, K_S$) are both fitted with a blackbody model. We noted that \mission{CFHT} $i$ band and the \mission{VISTA} $Y$ band fluxes significantly exceed the summation of the best-fitting optical and NIR models, suggesting that maybe there is another component contributes significantly in these two bands. We thus excluded these two bands in our fitting. The \mission{GALEX} NUV band was not included as the NUV band may lag behind the optical bands (see Section\,\ref{subsec:plateau}). The best-fit models, dashed line for the optical bands and dot-dashed line for the NIR bands, are overplotted in Figure\,\ref{fig:sed_fitting}. The solid lines in Figure\,\ref{fig:sed_fitting} show the summation of the two best-fitting blackbody components.

\begin{figure}
	\includegraphics[width=\columnwidth]{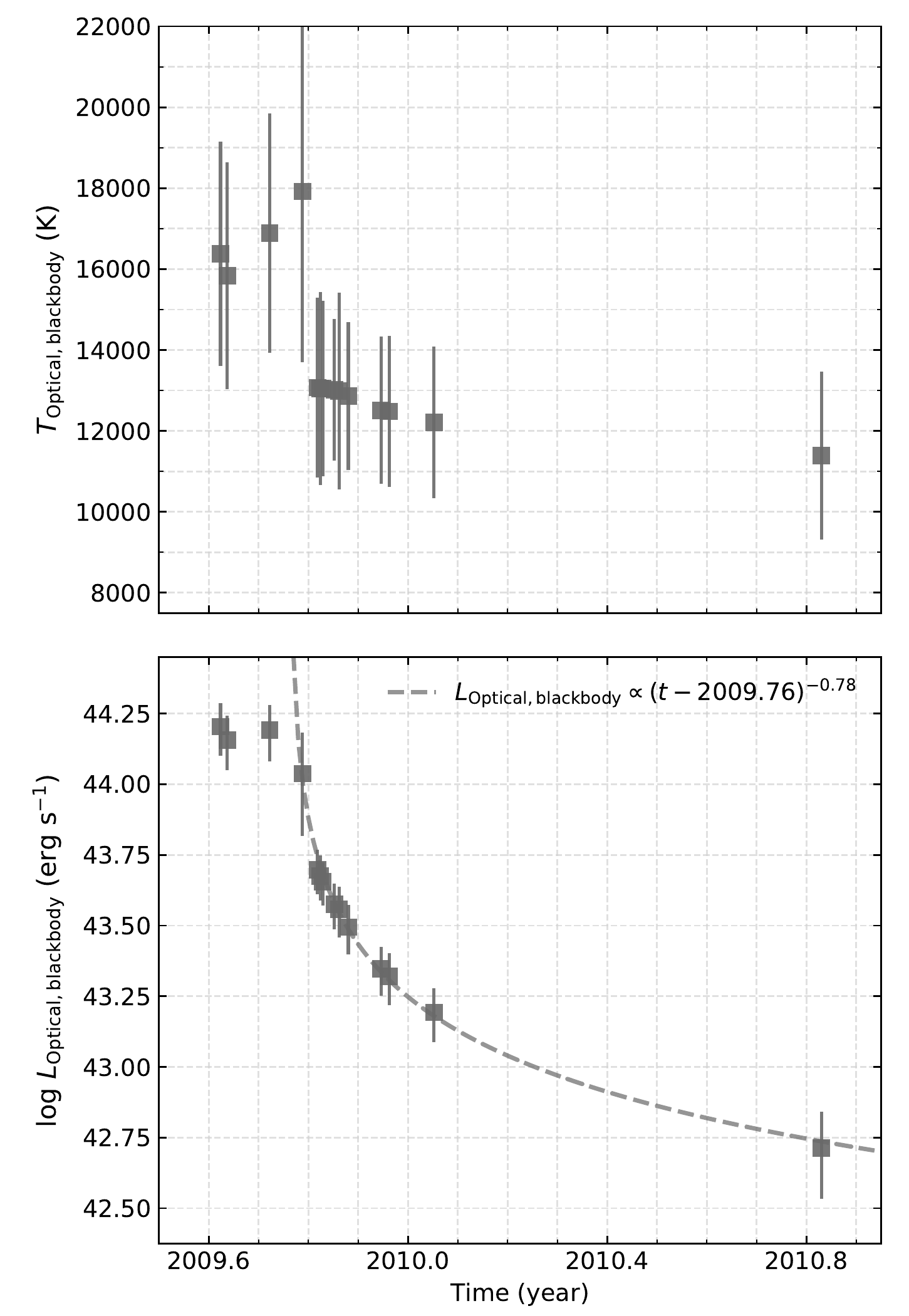}
	\caption{\label{fig:tbb_lbb}Upper panel: the disk temperature evolution of the blackbody component in optical bands; Bottom panel: the evolution of the luminosity of the blackbody component.}
\end{figure}

\begin{figure}
	\includegraphics[width=\columnwidth]{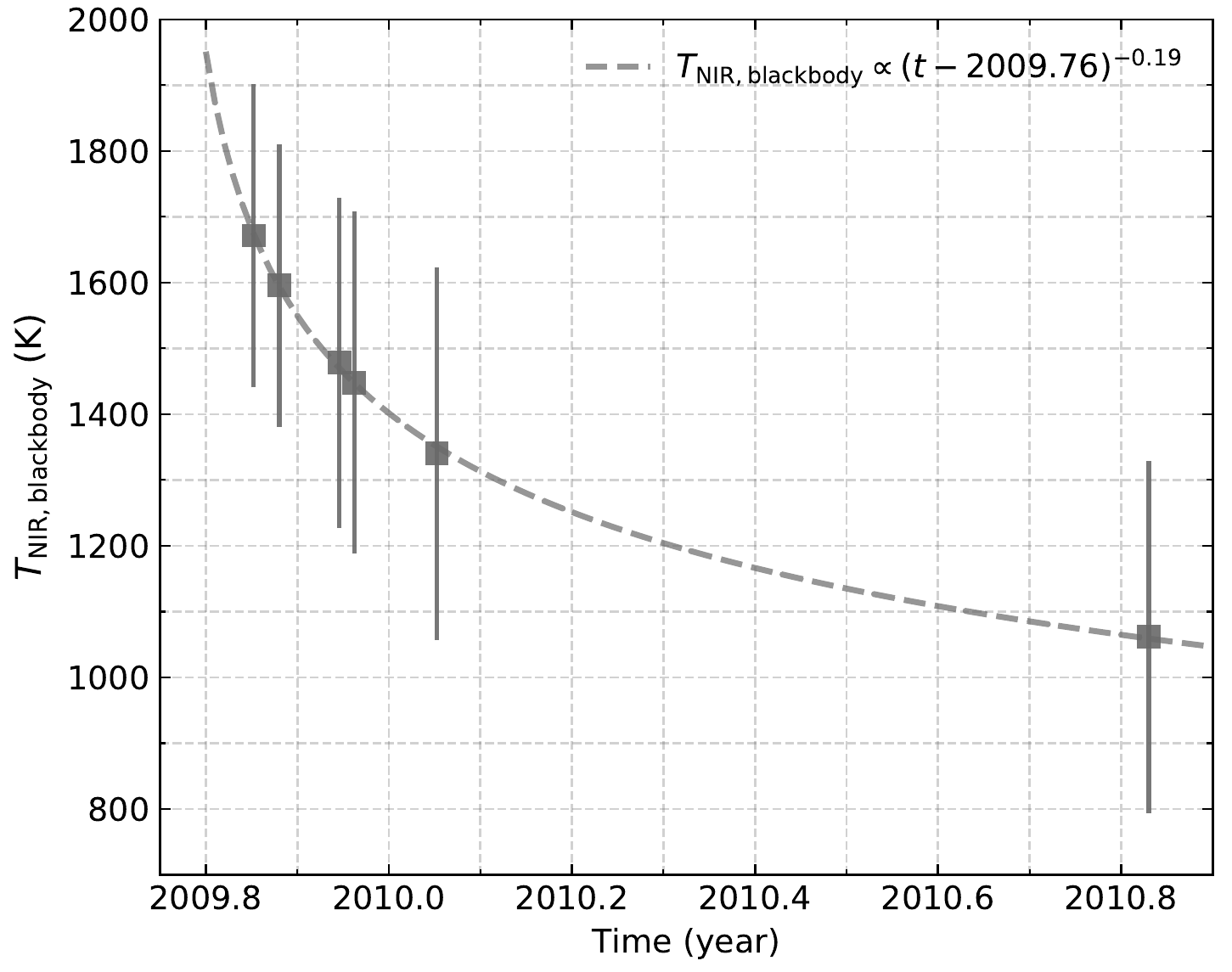}
	\caption{\label{fig:tbb_dust}The evolution of the blackbody temperature of the hypothesized dust component. The temperatures are measured with the \mission{VISTA} $J, H, K_\mathrm{S}$ bands.}
\end{figure}

In Figure\,\ref{fig:tbb_lbb}, we show the evolution of the optical blackbody temperature (upper panel) and bolometric luminosity (lower panel). Unlike the optical bolometric luminosity, which evolved following a power-law $L_{\mathrm{Optical, blackbody}}\propto(t-2009.77)^{-0.64}$ after the plateau phase, the temperature of the blackbody component did not vary significantly, in agreement with previous studies \citep[e.g.][]{blanchard_etal2017, wevers_etal2019}. The plateau phase can also be clearly seen in Figure\,\ref{fig:tbb_lbb}. The total energy released, calculated using the best-fitting optical blackbody model, from the TDE is estimated to be $\sim10^{51}\,\mathrm{erg}$ by integrating a constant luminosity function ($L_{\mathrm{Optical, blackbody}}\sim1.3\times10^{44}$\unitlumi) during the plateau phase and the best-fitting power-law function after the plateau phase. The estimated temperature of the NIR blackbody component at different time epochs can be seen in Figure\,\ref{fig:tbb_dust}. The temperatures, $T_{\mathrm{NIR, Blackbody}} \lesssim 1700\,\mathrm{K}$, are consistent with the expected temperature of thermal radiation from the dust. It evolved roughly following a power-law function, $T_{\mathrm{NIR, blackbody}}\propto(t-2009.76)^{-0.19}$.

%%%%%%%%%%%%%%%%%%%%%%% Discussion %%%%%%%%%%%%%%%%%%%%%

\section{Discussion\label{sec:discussion}}

\subsection{TDE, AGN flare or Supernovae?\label{subsec:tdeoragn}}
The X-ray spectral property and the detection of the broad optical emission lines clearly reveal the AGN nature of J0227-0420. It is known that AGNs are highly variable sources in almost the whole electromagnetic spectrum. Giant flares have also been detected in the IR-to-X-ray data of some AGNs. For instance, a large amplitude X-ray variability with time-scale of years was discovered in the low luminosity AGN NGC 7589 by \citet{yuan_etal2004}. The optical spectroscopic analysis and the X-ray spectral property suggest that the variability is likely caused by accretion state transition. A giant flare, which lasted for at least a month, is found in the multi-band data of NGC 1566 \citep{oknyansky_etal2017}. The X-ray spectral property during the flaring period of NGC 1566 suggested that the flare is likely caused by disk instability, rather than by a TDE \citep{parker_etal2019}. The highly variable AGN Mrk 335 also frequently shows UV flares which are also not caused by TDEs in the long-term \mission{Swift} monitoring data \citep{gallo_etal2018}. However, optical flares with duration longer than $\sim 3$ years and amplitude larger than $0.5$\,mag is rare in AGNs, e.g. 51 such flares are found out of more than 900000 AGNs \citep[$<6\times10^{-5}$,][]{graham_etal2017} in the Catalina Real-time Transient Survey. More importantly, most of these flares have a raising time-scale comparable to and even longer than the decline time-scale. While the flare discovered in J0227-0420 shows a relatively rapid raising time-scale (shorter than 6 months) and a decline time-scale longer than 3 years (see Figure\,\ref{fig:multi_lc}). In addition, the multi-band long-term light curves can all be well fitted with a power-law (after the plateau phase for optical and NUV bands, see Section\,\ref{subsec:modelling_lc}). These features are in agreement with the predicted characteristics of TDEs. We thus argue that the observed increase and long-term decline of the multi-band fluxes were caused by a TDE occurred in the NLS1 galaxy J0227-0420 (\citealt{blanchard_etal2017} reported a TDE candidate PS16dtm which was also occurred in a NLS1 galaxy), though we note that a giant AGN flare due to other processes than TDEs can not be ruled out.

A supernova which happened at the center of J0227-0420 may also produce a giant flare that has similar profile as the observed long-term evolution of J0227-0420. The observed plateau phase in the optical and UV bands then suggested a Type IIP supernova. However, the majority of Type II SNe have a total luminosity in the range $10^{41}-10^{43}$\unitlumi and total energy lower than $10^{50}$\,erg, i.e. at least an order of magnitude lower than the estimated luminosity/energy of J0227-0420. In addition, at the early stage, the temperature of the optical blackbody component of J0227-0420 is almost constant, which is also significant difference from that observed in Type II supernova. Moreover, contrary to the general decline trend of Type IIP supernova, the multi-band data of J0227-0420 showed a power-law rather than exponential decline. Thus we conclude that the flare observed in J0227-0420 should not be caused by a supernova event.

\subsection{The plateau phase: origin of UV/optical emission\label{subsec:plateau}}
Overall, the profile of the multi-band light curves of the flare are in agreement with a power-law form. However, a plateau phase, or a slowly decline phase, is evidently present in the CFHT $u, g, r, i$ and \mission{GALEX} NUV data. A plateau phase has been reported in a few TDEs in the literature, e.g. PS16dtm \citep[][claimed to be a TDE in a NLS1 galaxy]{blanchard_etal2017} and AT 2018fyk \citep{wevers_etal2019}. Similar to both PS16dtm and AT 2018fyk, the temperature, when measured with a simple blackbody model (using only the \mission{CFHT} $u, g, r$), is also almost constant during the plateau phase for J0227-0420. However, the plateau phase observed in J0227-0420 is different from those reported before, which lasted for $\sim 100\,$days for PS16dtm and $40$ days for AT 2018fyk. No apparent dependence of the duration of the plateau phase on wavelength (e.g. \mission{Swift/UVOT} $UVW1, UVW2, UVM2, U$ ) was found for both sources. In J0227-0420, the duration of the plateau phase is much shorter in the NUV band ($\sim7-15$ days) than in the optical bands ($\sim 40-50$ days). In addition, the plateau phase in the NUV band lagged behind that in the optical band for nearly $\sim80$ days. We caution here that there is no NUV observation at the time of the optical plateau phase, so the possibility that a plateau phase started at the same time with similar duration as the optical plateau phase can not be ruled out.

Based on the emergence of narrow emission lines, such as the low ionization Fe\,\textsc{ii} lines during the plateau phase,  \citet{wevers_etal2019} proposed that the plateau phase observed in the optical bands is ascribed to the reprocessing of the X-ray emission by the newly formed accretion disk. The decline of the optical/UV luminosity, before the plateau phase, is due to a shock caused by self-interactions between tidal streams. However, there should be no lag between the NUV and optical plateau phases in this scenario, which is inconsistent with the $\sim80$ days lag observed in J0227-0420.

We propose the following scenario to qualitatively explain the observed multi-band data of J0227-0420. The nodal precession introduced by the central spinning BH may change the orbital plane of the stellar debris streams. Thus the streams with different semi-major axes can have different orbital angles with respect to the BH equator with different semi-major axis. As was shown in the numerical simulations by \citet{chan_etal2019}, if the streams are dense enough (e.g. $\dot{M_\mathrm{s}}/\dot{M_\mathrm{d}}$ is large, where $\dot{M_\mathrm{s}}$ is the early time mass current of the debris stream, and $\dot{M_\mathrm{d}}$ is the mass current of the pre-existing disc at the interaction radius), only a fraction of the streams will be captured by the pre-existing disk during the collision between the streams and the disk. For J0227-0420 with BH mass of $\sim10^6\,\msun$ and Eddington ratio of $\sim0.6$, the $\dot{M_\mathrm{s}}/\dot{M_\mathrm{d}}$ is likely to be larger than 100 \citep[see figure 1 of][]{chan_etal2019}. Thus a large fraction of the material will penetrate through the disk. Part of those outgoing material will remain bound to the BH with a smaller specific binding energy and surface density. After one or several orbital periods, they will eventually again collide with the disk at relatively large distances, and are captured by the disk. This suggests that the stellar debris can be captured by the disk at very different radii, which can then explain the initial decline observed in the NUV/optical emission. On the other hand, due to the loss of the angular momentum, the material in the outer disk will gradually move inward and increase the local accretion rate at the inner region, producing the optical and UV plateau phases.
 
 The lag between the NUV and optical plateaus is then determined by the viscosity time-scale $\tau_{\mathrm{vis}} \sim R / v_{R}=R^{2} /\left(\alpha c_{\mathrm{s}} H\right)$ \citep{frank_jhan_king2002, kato_fukue_mineshige2008}, where $\alpha$ is the viscosity parameter; $v_{R}$ and $c_{\mathrm{s}}$ are the radial velocity and the sound speed at radius $R$, respectively; $H$ is the scale height of the accretion disc. Following \citep{cao_wang2014} the viscosity time-scale at a given (truncation) radius $R_{\mathrm{tr}}$ is estimated using the following equation: % which can be calculated by

\begin{equation}
\tau_{\text { vis }} \sim 1.56 \times 10^{-7} r_{\mathrm{tr}}^{3 / 2} \alpha^{-1}\left(\frac{H}{R}\right)^{-2}\left(\frac{M_{\mathrm{BH}}}{10^{6} \mathrm{M}_{\odot}}\right) \mathrm{yr,}
\end{equation} 
 where $r_{\mathrm{tr}}=R_{\mathrm{tr}}/R_{\mathrm{g}}$. The $\tau_{\text { vis }}$ strongly depends on the poorly known viscosity parameter $\alpha$ and the disk thickness $H/R$. For instance, $\tau_{\text { vis }}$ is more than several tens years at a radii of $\sim1500\,R_{\mathrm{g}}$ (optical radiation region) for a standard thin disk with $\alpha=0.3$, $H/R\sim0.01$, and $M_{\mathrm{BH}}=10^6\msun$, while it can be as short as several months for a thick disk with $H/R\sim1$. If the accretion disk in J0227-0440 is geometrically thick, which is favored by its high accretion rate close to the Eddington ratio, the observed lag of $\sim3$ months can be explained by the viscosity time-scale.
 
\subsection{Evolution of the light curves in different wave-bands: implications for reprocessing in NIR and MIR\label{subsec:band_vs_lc}}
The index of the light curves exhibits a dependence on wavelength. On the other hand, the indices are consistent within uncertainties in a certain wave-bands, $\sim0.8-0.9$ in the optical bands and $\sim1.2$ in the MIR bands, while a steeper index with value larger than 2.0 in the NIR bands {(\bf  except for the $Y$ band which has an index less than 2.0. This may partly due to the contribution from the optical blackbody component)}. This implies that the observed UV/optical, NIR, and MIR lights are radiated from different regions. The measured temperature of the blackbody component is more than $10^4$\,K using the optical data, in agreement with the predicted temperature of the outer region of the accretion disk\footnote{We note here that the temperature is estimated using the host and AGN subtracted data. In reality, the accretion material from the AGN and TDE are indistinguishable, and should be modeled together.}, though with a steeper decline rate than the prediction from some theoretical calculations \citep{lodato_rossi2011}.

MIR flares have been detected in TDEs happened in quiescent galaxy and AGNs \citep[e.g.][]{dou_etal2016, dou_etal2017, vanvelzen_etal2016, jiang_etal2016, jiang_etal2017, jiang_etal2019, wang_etal2018, sheng_etal2019}. They are normally explained as the dust echoes, i.e. dust reprocessing of the UV/optical radiation from TDE. Under this scenario, the distance of the dust material to the region of the primary UV/optical emission can be measured via the delay of the MIR to the UV/optical emission. Due to the lack of observations before and during the TDE in MIR and the non-detection of the peak radiation in all the wavelength bands, the delay time is difficult to be estimated for J0227-0420. Nevertheless, the best-fitting peak time $t_0$ (see Table\,\ref{tab:lc_fit}, note that the true peak time for the optical band should be earlier than the best-fitting value quoted in the table, as the plateau phase is not included in the fitting), though with large uncertainties, indicates that probably the MIR flare happened soon after the UV/optical flare, e.g. less than half a year. A small delay time ($\sim110$\,days) has actually been found in the dust echo in ASASSN-14li \citep{jiang_etal2016}. The dust sublimation radius of a typical AGN can be estimated by  \citep{namekata_umemura2016}
$$ R_{\mathrm{sub}} = 0.121\,\mathrm{pc}\,\bigg(\frac{L_\mathrm{bol}}{10^{45}\,\mathrm{erg\,s^{-1}}}\bigg)^{0.5}\bigg(\frac{T_\mathrm{sub}}{1800\,\mathrm{K}}\bigg)^{-2.804}\bigg(\frac{a}{0.11\,\mathrm{\mu m}}\bigg)^{-0.51},$$
where $L_\mathrm{bol}$ is the bolometric luminosity of the AGN, $T_\mathrm{sub}$ and $a$ are the sublimation radius and size of the dust grain, respectively. The dust sublimation radius is then about $\sim40$ light days for J0227-0420, assuming a sublimation temperature of 1850\,K and dust grain size of $0.1\,\mathrm{\mu m}$. Thus the delay between the MIR and optical/UV could be small.

The best-fitting temperature of the blackbody component, measured by fitting the NIR \textit{J, H, K$_\mathrm{S}$} bands, is consistent with the temperature of the dusty torus. In addition, the NIR flux are much higher than the values expected from the extrapolation of the optical/UV blackbody model. Those results imply that the NIR emission are also likely dominated by the radiation from the dusty torus. The decline rates in the NIR $J, H, K_\mathrm{S}$ bands, however, are much faster than the optical ones, with indices $\gtrsim 2.0$. While in the dust reprocessing scenario, a flatter index is expected due to the delay time between the central source and the dust torus, i.e. $L_{\mathrm{IR}}(t)=\int{\psi(\tau)L_{\mathrm{Primary}}(t-\tau)\,\mathrm{d}\tau}$, where $\tau$ is the delay time, $\psi(\tau)$ is the response function, and $L_{\mathrm{Primary}}$ and $L_{\mathrm{IR}}(t)$ are the light curves of the primary optical/UV and reprocessed IR emission, respectively.

One possible explanation for the steeper indices may be that the dust grains in the inner region of the torus are evaporated during the strong flare phase. For a given incident flux and emissivity, the IR luminosity for dust grains with size $a$ will depend on the number density $n(a)$ as well as the surface area of the dust grain. In this case, the response function $\psi(\tau)$ is a function of not only the delay time $\tau$ but also the variability time scale $\tau\prime$ of the IR luminosity caused by the evaporation of dust grains. The flux increase in the NUV and optical \textit{u} band may effectively heat up and potentially evaporate the dust grains, especially these small size dust grains which are normally hotter and may significantly contribute to the NIR radiation. To explain the steep decline rate with dust evaporation requires that the dust evaporation time-scale should not be too  long (the variability should be flatter than that of the primary emission), comparing to the time scale of the flare (full width at half maximum larger than $3$\,months) and also the time decay due to the finite speed of light and the geometrical extent of the dusty torus ($\tau_{\mathrm{decay}}\sim2R_{\mathrm{sub}}/c\approx3$\,months). For graphite dust grain, the grain survive time can be given as \citep{vanvelzen_etal2016},
$$\begin{aligned} t_{\text { sub }} &=\frac{a}{\mathrm{d} a / \mathrm{d} t} \\
& \simeq 3.21 a_{0.1} \exp \left[42.747\left(1900 / T_{\mathrm{d}}-1\right)\right] \text { month }, \end{aligned}$$
where $a_{0.1}$ is the grain size in unit of $0.1\micron$, $T_{\mathrm{d}}$ is the sublimation temperature. Assuming a sublimation temperature of 1850 K, the time-scale for grain with size of $0.1\micron$ will be $\sim10$ months. Smaller size dust grains will have much shorter sublimation timescale, e.g. $t_{\text {sub}} \sim 5$ months for grain size of $0.05\micron$, which could be comparable to the time-scale of TDE and $\tau_{\mathrm{decay}}$, suggesting that $t_{\text { sub }}$ may have significant impact on the response function. Thus it is possible that the sublimation of dust could contribute to the rapid decline of the NIR radiation. Future detailed theoretical self-consistent calculation is necessary to confirm the effect of the dust sublimation.

\section{Summary\label{sec:summary}}

In this work, we report the discovery of a TDE candidate in the AGN SDSS J0227-0420. The AGN nature of J0227-0420 is revealed by the persistent X-ray radiation and the detection of the optical broad emission lines before the flare. A large amplitude flare, lasting for more than 3 years, is clearly shown in the long-term IR/optical/UV light curves. The location of the flare is in agreement with the position of the AGN, suggesting an emission region close to the central BH. The long-term multi-bands light curves can be well described with a power-law, consistent with the expectation from a TDE. A plateau phase is evidently seen in the NUV and optical data. Moreover, the NUV plateau may lag behind the optical ones by $\sim70-80$ days with also a much shorter duration ($\sim15$ days in NUV and $\sim30-40$ days in the optical). The sudden increase in the multi-band fluxes can be attributed to a TDE in J0227-0420. The stellar debris collided with the pre-existing disk at different disk radii, producing the initial increase in the flux. The material in the outer disk then gradually drifted toward the BH and increased the local accretion rate at inner disk radii, resulting in the plateau phases observed in the optical and NUV bands with the observed time lag. In this scenario, the lag between the NUV and optical plateau phase is then due to the viscosity decay. Flares in the NIR and MIR bands, which can be explained as the dust echoes of the primary flare, are also seen in J0227-0420.  The NIR flux showed a much faster decline rate than that in the optical bands, which may be explained as that the dust grains in the inner region are sublimated by the intense radiation of the TDE flare. Our results imply that distinctive features may be present in TDEs occurring in AGNs, which can be helpful in understanding the accretion process. X-ray missions with large field of view, such as eROSITA \citep{merloni_etal2012} and the future Chinese X-ray satellite Einstein Probe \citep{yuan_etal2016}, will discover more TDEs in both quiescent galaxies and AGNs.

\section*{Acknowledgements}

ZL thanks Professor Xinwu Cao, Dr. Ning Jiang, Dr. Chichuan Jin, and Dr. Erlin Qiao for helpful discussions and comments. This work is supported by the Strategic Pioneer Program on Space Science, Chinese Academy of Sciences, grant No. XDA15052100. WY  acknowledges the support from the Strategic Priority Research Program of the Chinese Academy of Sciences grant No. XDB23040100. This work is also supported by the National Natural Science Foundation of China (grant no. 11673026, U1631238). This work is based on observation obtained with \mission{XMM-Newton}, an ESA science mission with instruments and contributions directly fund by ESA Member States and NASA, based on observations made with the NASA/ESA Hubble Space Telescope, and obtained from the Hubble Legacy Archive, which is a collaboration between the Space Telescope Science Institute (STScI/NASA), the Space Telescope European Coordinating Facility (ST-ECF/ESA) and the Canadian Astronomy Data Centre (CADC/NRC/CSA). The \mission{Chandra} data are obtained from the \mission{Chandra Data Archive}. This research has made use of software provided by the Chandra X-ray Center (CXC) in the application packages CIAO. Based on observations obtained with MegaPrime/MegaCam, a joint project of CFHT and CEA/IRFU, at the Canada-France-Hawaii Telescope (CFHT) which is operated by the National Research Council (NRC) of Canada, the Institut National des Science de l'Univers of the Centre National de la Recherche Scientifique (CNRS) of France, and the University of Hawaii. This work is based in part on data products produced at Terapix available at the Canadian Astronomy Data Centre as part of the Canada-France-Hawaii Telescope Legacy Survey, a collaborative project of NRC and CNRS. This publication makes use of data products from the Wide-field Infrared Survey Explorer, which is a joint project of the University of California, Los Angeles, and the Jet Propulsion Laboratory/California Institute of Technology, and NEOWISE, which is a project of the Jet Propulsion Laboratory/California Institute of Technology. WISE and NEOWISE are funded by the National Aeronautics and Space Administration.

\software{SAS (v16.1; \citealt{gabriel_etal2004}), gPhton \citep{million_etal2016b,million_etal2016a}, Elixir pipeline (v2; \citealt{magnier_cuillandre2004}), SCAMP (v2.7.1; \citealt{bertin_2006}), SExtractor (v2.25.0; \citealt{bertin_arnouts1996}), Xspec (v12.10; \citealt{arnaud1996}), CIAO (v4.10; \citealt{fruscione_etal06}), emcee \citep{foreman-mackey_etal2013}, lmfit (\citealt{newville_etal2018}, https://doi.org/10.5281/zenodo.1699739), Astropy \citep{astropy}, Matplotlib \citep{matplotlib}, Numpy \citep{numpy}, Scipy \citep{scipy}}

%%%%%%%%%%%%%%%%%%%% REFERENCES %%%%%%%%%%%%%%%%%%

% The best way to enter references is to use BibTeX:

\bibliography{./references}{}
\bibliographystyle{aasjournal}

% Alternatively you could enter them by hand, like this:
% This method is tedious and prone to error if you have lots of references
%\begin{thebibliography}{99}
%\end{thebibliography}

%%%%%%%%%%%%%%%%% APPENDICES %%%%%%%%%%%%%%%%%%%%%

%\appendix

%\section{Some extra material}

%%%%%%%%%%%%%%%%%%%%%%%%%%%%%%%%%%%%%%%%%%%%%%%%%%

\end{document}

%% file: xray_spec_fit_info.tex
\begin{table*}
%\begin{center}
\caption{\label{tab:xray_fitting}X-ray observation log and the best-fitting parameters for the X-ray spectra.}
\begin{tabular}{cccccccccc}
\hline\hline
  {Obs. ID} & {Obs. Date} & {Count rate\tablenotemark{a}} & {$T_\mathrm{BB}$} & {$\log f_\mathrm{Powerlaw}$} & {$\Gamma$} & {$E_\mathrm{Gaussian}$} & {\textit{EW}} & {$L_\mathrm{0.3-10\,keV}$} \\ %\multicolumn{1}{c}{$L_\mathrm{2-10\,keV}$} \\
  
             &  & pn/M1/M2 & keV & \unitflux & & keV & eV & \unitlumi\\\hline % & \unitlumi \\\hline
  0112680101 & 2002-01-28 & 0.41/0.10/0.10 & $ 0.13_{-0.01}^{+0.01} $ & $ -11.95_{-0.02}^{+0.03}$ & $2.39_{-0.09}^{+0.08}$ &  $6.37_{-0.36}^{+0.23}$ & $435^{+363}_{-208}$ & 8.50E42 \\[1mm] % & 2.41E42
  0112681301 & 2002-07-26 & 0.43/0.11/0.10 & $ 0.14_{-0.01}^{+0.01} $ & $ -11.94_{-0.03}^{+0.03}$ & $2.36_{-0.11}^{+0.11}$ &   --- &   ---  &  8.48E42 \\[1mm] %& 2.46E42
  0780451501 & 2017-01-09 & 0.47/0.10/0.09 & $ 0.14_{-0.01}^{+0.01} $ & $ -11.99_{-0.07}^{+0.07}$ & $2.16_{-0.28}^{+0.26}$ &   --- &   ---  &  9.12E42 \\[1mm] %& 2.75E42
  0780451601 & 2017-01-09 & ---/---/0.06   & $ 0.15_{-0.01}^{+0.01} $ & $ -12.04_{-0.05}^{+0.05}$ & $1.92_{-0.29}^{+0.27}$ &  $6.51_{-1.11}^{+0.34}$ & $705^{+1638}_{-477}$ & 8.65E42 \\[1mm]%& 3.25E42
  0780451701 & 2017-01-10 & 0.60/0.15/0.12 & $ 0.13_{-0.01}^{+0.01} $ & $ -11.82_{-0.03}^{+0.03}$ & $2.37_{-0.10}^{+0.09}$ &  $6.58_{-0.10}^{+0.21}$ & $389^{+544}_{-210}$  & 1.20E43 \\[1mm]%& 3.29E42
  0780451801 & 2017-01-10 & 0.82/0.18/0.16 & $ 0.13_{-0.01}^{+0.01} $ & $ -11.88_{-0.02}^{+0.03}$ & $2.38_{-0.08}^{+0.08}$ &  $6.40_{-0.54}^{+0.29}$ & $235^{+280}_{-135}$  & 1.02E43 \\[1mm]%& 2.88E42
\hline
\end{tabular}
%\end{center}
{\hspace*{2.1cm}$^a$Background subtracted}
\end{table*}
%
%$ -12.59_{-0.12}^{+0.09} $ & 
%$ -12.63_{-0.19}^{+0.12} $ & 
%$ -12.34_{-0.23}^{+0.13} $ & 
%$ -12.35_{-0.17}^{+0.10} $ & 
%$ -12.38_{-0.12}^{+0.09} $ & 
%$ -12.49_{-0.12}^{+0.09} $ & 

%% file: lc_fit_results.tex
\begin{table*}
%\begin{center}
\caption{\label{tab:lc_fit}Best-fitting parameters for the multi-bands light curves.}
\begin{tabular}{lcccccc}
\hline\hline
  \multicolumn{1}{l}{Band} &
  \multicolumn{1}{l}{Central wavelength\tablenotemark{a}} &
  \multicolumn{1}{l}{Band width\tablenotemark{a}} &
  \multicolumn{1}{c}{$\log{A}$} &
  \multicolumn{1}{c}{$t_0$} &
  \multicolumn{1}{c}{$\beta$} &
  \multicolumn{1}{c}{Constant}\\
  \multicolumn{1}{c}{} &
  \multicolumn{1}{c}{\AA} &
  \multicolumn{1}{c}{\AA} &
  \multicolumn{1}{c}{\unitflux} &
  \multicolumn{1}{c}{year} &
  \multicolumn{1}{c}{} &
  \multicolumn{1}{c}{$10^{-12}$\unitflux}\\
\hline
  CFHT $u$    & 3554.6 & 754   & $-12.27\pm0.16$ & $2009.70\pm0.06$ & $-0.94\pm0.33$ &  ---\\
  CFHT $g$    & 4626.8 & 1447  & $-12.40\pm0.17$ & $2009.70\pm0.05$ & $-0.96\pm0.33$ &  ---\\
  CFHT $r$    & 5965.8 & 1213  & $-12.40\pm0.17$ & $2009.70\pm0.05$ & $-0.80\pm0.33$ &  ---\\
  CFHT $i$    & 7259.3 & 1595  & $-12.33\pm0.05$ & $2009.72\pm0.02$ & $-0.79\pm0.06$ &  ---\\[1mm]
  VISTA $Y$   & 9686.6 & 994   & $-12.72\pm0.53$ & $2009.56\pm0.22$ & $-1.65\pm0.66$ & $4.17\pm1.25$\\
  VISTA $J$   & 11871  & 1709  & $-13.13\pm0.20$ & $2009.59\pm0.07$ & $-2.67\pm0.61$ & $4.08\pm0.13$\\
  VISTA $H$   & 15670  & 2849  & $-12.69\pm0.26$ & $2009.60\pm0.10$ & $-2.14\pm0.61$ & $3.59\pm0.85$\\
  VISTA $K_S$ & 20417  & 2849  & $-12.24\pm0.47$ & $2009.60\pm0.14$ & $-2.07\pm0.64$ & $2.74\pm0.25$\\[1mm]
  WISE W1     & 32887  &   & $-12.03\pm0.27$ & $2009.66\pm0.09$ & $-1.36\pm0.41$ & $1.79\pm0.56$\\
  WISE W2     & 43685  &   & $-11.85\pm0.44$ & $2009.59\pm0.10$ & $-1.17\pm0.29$ & $1.45\pm0.22$\\
\hline
\end{tabular}
%\end{center}
{\hspace*{2cm}$^a$In the source rest-frame}
\end{table*}